# Effects of magnetic doping and temperature dependence on phonon dynamics in CaFe$_{1-x}$Co$_x$AsF compounds ($x$ = 0, 0.06, 0.12)

R. Mittal[1,2], M. Zbiri[3], S. Rols[3], Y. Su[1], Y. Xiao[4], H. Schober[3], S. L. Chaplot[2], M. Johnson[3], T. Chatterji[5], S. Matsuishi[6], H. Hosono[6] and Th. Brueckel[1,4]

[1]*Juelich Centre for Neutron Science, IFF, Forschungszentrum Juelich, Outstation at FRM II, Lichtenbergstr. 1, D-85747 Garching, Germany*
[2]*Solid State Physics Division, Bhabha Atomic Research Centre, Trombay, Mumbai 400 085, India*
[3]*Institut Laue-Langevin, BP 156, 38042 Grenoble Cedex 9, France*
[4]*Institut fuer Festkoerperforschung, Forschungszentrum Juelich, D-52425 Juelich, Germany*
[5]*Juelich Centre for Neutron Science, Forschungszentrum Juelich, Outstation at Institut Laue-Langevin, BP 156, 38042 Grenoble Cedex 9, France*
[6]*Frontier Research Center, Tokyo Institute of Technology, 4259 Nagatsuta-cho, Midori-ku, Yokohama 226-8503, Japan*

We report detailed measurements of composition as well as temperature dependence of the phonon density-of-states in a new series of FeAs compounds with composition CaFe$_{1-x}$Co$_x$AsF ($x$ = 0, 0.06, 0.12). The electronic structure calculations for these compounds show that bands near the Fermi level are mainly formed by Fe 3d states, which is quite different from other 122 and 1111 FeAs compounds, where both Fe and As are believed to be related to superconductivity. The difference in electronic structure for fluorine based compounds may cause phonon spectra to behave differently as a function of composition and temperature in comparison with our previous phonon studies on parent and superconducting MFe$_2$As$_2$ (M=Ba, Ca, Sr). The composition as well as temperature dependence of phonon spectra for CaFe$_{1-x}$Co$_x$AsF ($x$ = 0, 0.06, 0.12) compounds have been measured using time of flight IN4C and IN6 spectrometers at ILL, France. The comparison of phonon spectra at 300 K in these compounds shows that acoustic phonon modes up to 12 meV harden in the doped compounds in comparison to the parent CaFeAsF. While intermediate energy phonon modes from 15 meV to 25 meV are also found to shift towards high energies only in the 12 % Co doped CaFeAsF compound. The experimental results for CaFe$_{1-x}$Co$_x$AsF ($x$ = 0, 0.06, 0.12) are quite different from our previous phonon studies on parent and superconducting MFe$_2$As$_2$ (M=Ba, Ca, Sr) where low-energy acoustic phonon modes do not react with doping, while the phonon spectra in the intermediate range from 15 to 25 K are found to soften in these compounds. We argue that stronger spin phonon interaction play an important role for the emergence of superconductivity in these compounds. The lattice dynamics of CaFe$_{1-x}$Co$_x$AsF ($x$ = 0, 0.06, 0.12) compounds is also investigated using the ab-initio as well as shell model phonon calculations. We show that the nature of the interaction between the Ca and the Fe-As layers in CaFeAsF compounds is quite different compared with our previous studies on $CaFe_2As_2$.

The discovery of superconductivity in La based FeAs ($T_c$=26 K) compound has stimulated enormous interest [1-28] in the field of condensed matter physics. There are mainly three types of FeAs-based superconductors such as RFeAsO (R: Rare earth elements), $AFe_2As_2$ (A: alkaline-earth elements), and LiFeAs. So far, the highest Tc of 56 K has been found in $Gd_{1-x}Th_xOFeAs$ [5]. The structural, magnetic, electronic properties of all the compounds have been extensively investigated to understand the mechanism of superconductivity. In particular for all these compounds strong anomalies have been found [1-5] in the specific heat, resistivity and magnetic susceptibility in the temperature range of 110 to 180 K. These anomalies are now known to be a prerequisite for superconductivity in FeAs compounds. In addition, it should be noted that recent results suggest that superconductivity and magnetism coexists [6,7] in these compounds.

Research groups are continuously making efforts to synthesize new FeAs compounds with higher $T_c$. Very recently new oxygen-free FeAs compounds MFeAsF (*M* = Sr, Ca and Eu) with ZrCuSiAs-like phase have been synthesized [9-14]. These new compounds are analogous to RFeAsO, where the (RO)$^+$ layer is replaced by (MF)$^+$ layer. Spin density wave anomalies have been found [12] in CaFeAsF and SrFeAsF at about 118 K and 180 K, respectively. The bulk superconductivity of 56 K and 32 K has been found [10] in SrFeAsF on partial substitution of Sr by Sm and La respectively. For CaFeAsF, the Co and Ni doping on the Fe site induces the superconductivity [13] with a *$T_c$* of 22 K and 12 K, respectively, while the discovery of superconductivity has been reported [14] in $Ca_{0.4}R_{0.6}FeAsF$ (R=Nd, Pr) with $T_c$ of 57.4 K and 52.8 K by doping of Nd and Pr, respectively. Neutron powder diffraction experiments have been carried out [15] to investigate the structural phase transition as well as magnetic order in $CaFe_{1-x}Co_xAsF$ compounds (*x* = 0, 0.06, 0.12). In this context, the relation between superconductivity, magnetic behaviour and structural instabilities have been studied. Resonant spin excitations have been observed [16(b)] in a number of FeAs compounds by inelastic neutron scattering. Both inelastic neutron scattering and first principle electronic band structure calculations indicate that superconductivity in the FeAs-based compounds is unconventional and might be mediated by magnetic spin fluctuations [17]. The suppression of the magnetic ordering giving rise to superconductivity requires lattice excitations leading to a coupling of the magnetic and structural degrees of freedom.

Hence, it is important to experimentally measure the phonon excitations and search for specific features in the phonon spectra. Raman spectra [18] of non superconducting $CaFe_2As_2$, $SrFe_2As_2$ and superconducting $Sr_{0.6}K_{0.4}Fe_2As_2$ have been measured.

Inelastic x-ray scattering has been used to investigate [19] the phonon density of states in $LaFeAsO_{1-x}F_x$ and $NdFeAsO$. Only very few phonon studies have been reported [19(a), 20] for single crystals of $BaFe_2As_2$ and $PrFeAsO_{1-y}$ using inelastic x-ray scattering. Ab-initio phonon calculations and shell model calculations have been used to calculate the phonon spectrum of these compounds. We have also investigated these compounds using the technique of inelastic neutron scattering from polycrystalline samples [21-24] of $BaFe_2As_2$, $CaFe_2As_2$, $Sr_{0.6}K_{0.4}Fe_2As_2$ and $Ca_{0.6}Na_{0.4}Fe_2As_2$ and a single crystal [25] of $CaFe_2As_2$.

The valence-band electrons, close to the Fermi surface, are mainly believed to be involved in the superconductivity. The electronic structure calculation for MFeAsF shows that bands near the Fermi level are mainly formed by Fe 3d states [26]. The Fe partial density of states at the Fermi level increases on going from SrFeAsF to CaFeAsF. This is quite different from RFeAsO (R: Rare earth elements) and $MFe_2As_2$ (M: Alkaline-earth elements) compounds where bonding and interaction of both Fe and As are believed to be responsible for superconductivity. We are conducting systematic studies of the temperature as well as composition dependence of phonon spectra for FeAs compounds. It is of interest to investigate the phonon spectrum of FeAs compounds as a function of increase in dopant concentration. In this continuation of our studies on FeAs compounds we now report the experimental temperature dependence of phonon spectra for $CaFe_{1-x}Co_xAsF$ ($x$ = 0, 0.06, 0.12) compounds. Phonons have also been estimated by means of lattice dynamics calculations in order to analyze the observed spectra. The details about the experimental and lattice dynamical techniques are given in Section II and Section III, respectively. The obtained results are given in Section IV. Finally Section V is dedicated to the conclusions.

**II. Experimental**

The polycrystalline samples of CaFeAsF, $CaFe_{0.94}Co_{0.06}AsF$ (onset $T_c$=20 K) and $CaFe_{0.88}Co_{0.12}AsF$ ($T_c$=20 K) were prepared by solid-state synthesis techniques [12]. All the samples were extensively characterized [15] by means of neutron powder diffraction, dc magnetization and electrical resistivity and confirmed their high quality. The neutron powder diffraction measurements were performed on the high flux powder diffractometer D20 at Institut Laue Langevin (Grenoble,

France). Rietveld refinement of the diffraction data indicates that samples contain small impurity phases (CaF$_2$ and Fe$_2$O$_3$) of less than 1%. The details about the characterization of our samples are given in our previous publication [15].

The inelastic neutron scattering experiments were performed using the IN4C and IN6 time of flight spectrometers at the Institut Laue Langevin (ILL), France. An incident neutron wavelength of 1.1 Å (67.6 meV) was chosen for the IN4C measurements, which allowed the data collection in the neutron-energy loss mode. The measurements for CaFe$_{1-x}$Co$_x$AsF compounds ($x$ = 0, 0.06, 0.12) using the IN4C spectrometer were carried out at 2 K, 125 K and 150 K. The tetragonal to orthorhombic phase transition in CaFe$_{0.94}$Co$_{0.06}$AsF is determined to be at around 85(3) K. So for the $x=0.06$ compound we have collected one more data set at 50 K. The high-resolution data for all the three compounds at 300 K were also collected using IN6. For these measurements we have used an incident neutron wavelength of 5.1 Å (3.12 meV). The full energy range of the phonon spectra at IN6 can only be covered by performing measurements in the neutron energy gain mode. In the incoherent one-phonon approximation the measured scattering function $S(Q,E)$, as observed in the neutron experiments, is related to the phonon density of states [29] as follows:

$$g^{(n)}(E) = A < \frac{e^{2W_k(Q)}}{Q^2} \frac{E}{n(E,T) + \frac{1}{2} \pm \frac{1}{2}} S(Q,E) > \qquad (1)$$

$$g^n(E) = B \sum_k \{\frac{4\pi b_k^2}{m_k}\} g_k(E) \qquad (2)$$

where the + or – signs correspond to energy loss or gain of the neutrons respectively and where $n(E,T) = [\exp(E/k_BT) - 1]^{-1}$. $A$ and $B$ are normalization constants and $b_k$, $m_k$, and $g_k(E)$ are, respectively, the neutron scattering length, mass, and partial density of states of the $k^{th}$ atom in the unit cell. The quantity between < > represents suitable average over all $Q$ values at a given energy. $2W(Q)$ is the Debye-Waller factor. The weighting factors $\frac{4\pi b_k^2}{m_k}$ for various atoms in the units of barns/amu are: Ca: 0.071; Fe: 0.208 and As: 0.073; F: 0.211; Co: 0.095. The values of neutron scattering lengths for various atoms can be found from Ref. [30]. The experimental phonon data is always contaminated by the multi-phonon contribution. In order to compare the experimental data with the calculated one-phonon spectra one has to subtract the multi-phonon contribution from the measured phonon spectra at each temperature. We have calculated multi-phonon contribution by Sjolander formalism [31] and subtracted it from the experimental data.

**III. Lattice dynamical calculations**

We have measured temperature dependence of phonon spectra for $CaFe_{1-x}Co_xAsF$ (x = 0, 0.06, 0.12) compounds. The calculation of phonon spectra are needed for interpretation of the experimental data. For CaFeAsF we have made an attempt to develop an interatomic potential model. Further, density functional calculations (DFT) have been undertaken to derive the phonon frequencies.

The lattice dynamical calculation are performed [32,33] using the following interatomic potentials

$$V(r) = \{\frac{e^2}{4\pi\varepsilon_o}\}\{\frac{Z(k)Z(k')}{r}\} + a\exp\{\frac{-br}{R(k)+R(k')}\} - \frac{C}{r^6} - D\exp[-n(r-r_o)^2/(2r)] \quad (3)$$

where $r$ is the distance between the atoms $k$ and $k'$. The first term is the long range Coulombic attractive potential, the second is the Born– Mayer repulsion and the third is the van der Waals attraction potential. The parameters of the interatomic potential are the effective charge Z(K) and radius R(k) of the atom type k. $1/(4\pi\varepsilon_o)$ = 9 x $10^9$ Nm$^2$/Coul$^2$, a=1822 eV, b=12.364. The third term in eq.(3) is applied only between Fe-As and As-As atoms. The radii parameters used in our calculations are R(Ca)= 2.18 Å, R(Fe/Co)= 0.42 Å, R(As)= 2.59 Å and R(F)=1.22 Å. Partial charges of Z(Ca)=2.00, Z(Fe/Co)=0.30, Z(As)= -1.35 and Z(F)= -0.95 are used in the calculations. The parameter *C* was set to C(Fe-As)= 70 eV/Å$^6$ and C(F-F)= 107 eV/Å$^6$. The covalent nature of the Fe-As bond has been described by further including the stretching potential given by the fourth term in Eq. (3). The parameters of the stretching potential are D=2.40 eV, n=9.2 Å$^{-1}$, $r_o$ =2.395 Å.

The polarizibility of the As and F atoms are introduced in the framework of the shell model [34] with the shell charge Y(As)=-1.45, Y(F)= -1.00 and shell-core force constant K(As)=40 eV/Å$^2$, K(F)=150 eV/Å$^2$. Due to the metallic nature of FeAs compounds, the screened Coulomb potential in the Thomas-Fermi approximation was used for the calculations. The screening parameter $k_o$=0.35 Å$^{-1}$ has been used in our calculations. The code [35] "DISPR" developed at Trombay is used for the calculation of phonon frequencies and polarization vector of the phonons in the entire Brillouin zone.

Further, DFT-based first principle calculations were performed using the projector-augmented wave (PAW) formalism [36] of the Kohn-Sham DFT [37, 38] at the generalized gradient approximation level (GGA), implemented in the Vienna ab initio simulation package (VASP) [39, 40]. The GGA was formulated by the Perdew-Burke-Ernzerhof (PBE) [41, 42] density functional. The Gaussian broadening technique was adopted and all results are well converged with respect to k-mesh and energy cutoff for the plane wave expansion. Experimentally refined crystallographic data in the low-temperature has been considered. This structure was used to calculate the GDOS and dispersion relations for the base-centered orthorhombic phase under the Cmma space group (number 67) having the local point group symmetry $D_{2h}^{21}$. In case of CaFeAsF, the Fe moments are parallel to the longer *a* axis. Spins are aligned ferromagnetically along the shorter *b* axis and antiferromagnetically along the *a* and *c* axes. In order to take into account computationally the observed magnetic ordering, one has to double the cell along the c-axis to get two layers of Fe cations and hence an antiferromagnetic ordering along this direction. This is specific to the present system when comparing to the other FeAs-based compounds, having two Fe-layers without doubling along c-axis. In the lattice dynamics calculations, in order to determine all inter-atomic force constants, the supercell approach has been adopted [43]. Therefore, the single cell was used to construct a (2*a, 2*b, 2*c) super cell containing 32 formula-units (128 atoms), a and b being the shorter cell axes. Total energies and inter-atomic forces were calculated for the 24 structures resulting from individual displacements of the four symmetry inequivalent atoms, along the three inequivalent Cartesian directions (±x, ±y and ±z). The 24 phonons branches corresponding to the 8 atoms in the primitive cell, were extracted in subsequent calculations using the Phonon software [44]. In the followings magnetic means that the supercell has been doubled along the c-axis (2×2×2) to get two Fe-layers (128 atoms), whereas non-magnetic phonons refer to calculations without doubling along the c-axis (super cell, 2×2×1, 64 atoms), i.e. no magnetism is considered.

By remaining in space group 67, the magnetic calculation described above results in a averaging of the magnetic interactions in the dynamical matrix constructed in Phonon software [44]. In order to include the precise ferro (b-direction) and antiferromagnetic (a and c directions) ordering in the dynamical matrix the space group symmetry has to be changed and a second Fe site has been introduced. Antiferromagnetic ordering in the a-direction reduces the symmetry to Pmma (49) but including the ordering in the c-direction raises the symmetry to Ibam (72), which is associated with a doubling of the unit cell. We refer to this more detailed magnetic calculation as the "broken symmetry" approach.

## IV. Results and discussion

### A. Comparison of phonon spectra for CaFe$_{1-x}$Co$_x$AsF compounds ($x$ = 0, 0.06, 0.12)

The high resolution density of states measurements for CaFe$_{1-x}$Co$_x$AsF compounds ($x$ = 0, 0.06, 0.12) measured at 300 K using the IN6 spectrometer are shown in Figure 1. The structural and magnetic transition in these compounds is below 134 K. The measurements on IN6 can only be performed with a small incident neutron energy of 3.12 meV in the neutron-energy gain mode, which does not give enough intensity at low temperatures. We have carried out density of states measurements only at 300 K. These measurements, carried out with very high elastic energy resolution of about 200 μeV (inelastic focusing mode), show that low-energy phonon modes below 12 meV harden on doping of 6 % of Co at the Fe site (Fig. 1). A further increase of the Co concentration to 12 % does not seem to affect the low energy phonon spectra. However, the intermediate energy phonon modes from 18 meV to 25 meV also harden in addition to the acoustic modes on 12 % Co doping in comparison to the parent and 6 % Co doped compounds.

The calculated partial density of states (Fig. 2) of various atoms (Section IVC) in CaFe$_{1-x}$Co$_x$AsF compounds ($x$ = 0, 0.06, 0.12) show that at low energies below 12 meV the contribution to phonon spectra is mainly from the Fe$_{1-x}$Co$_x$ or As sub lattice. The shell model calculation (Section IVC) which does not consider the effect of electron-phonon coupling influence on the phonon spectra shows that the partial substitution of Co at the Fe site has little effect (Figs. 1 and 3) on the phonon spectra.

Electronic structure calculations [26] show that for CaFeAsF bands near the Fermi level are mainly formed by Fe 3d states. The substitutions of Co at the Fe site would result in an increase in electrons in the doped compound in comparison to the parent compound. Indeed there are seven electrons in the d-shell of the Co$^{+2}$ cation, whereas the valence d-shell of Fe$^{+2}$ consists of six electrons. Thus partial substitution of Co by Fe would be reflected by an increase in electronic density of states of the Fe$_{1-x}$Co$_x$ sub-lattice at the Fermi level. The changes in the electronic system upon doping may be responsible for the hardening of the low-energy modes. The hardening can thus be taken as an indication for coupling between electrons and phonons (Fig. 1). Further increase in Co doping to 12 % would further perturb the electronic density of states of Fe$_{1-x}$Co$_x$ at the Fermi level. The additional increase in strength of the electronic density of states may cause intermediate energy phonons to interact with the electrons and in turn results in hardening of intermediate energy modes (Fig. 1).

The peak in the phonon spectra resulting from the stretching vibrations of Fe-As modes remains centered at about 32 meV in all the CaFe$_{1-x}$Co$_x$AsF compounds ($x$ = 0, 0.06, 0.12). This indicates that Fe-As bond lengths are not affected by partial substitution of Fe by Co atoms. This is what we also observe from the analysis of neutron diffraction [15] on these compounds. However, the peak at about 32 meV is found to be slightly broader in Co doped compounds in comparison of the parent compound. The slight distribution of Fe/Co-As bond lengths would broaden the 32 meV peak in Co doped compounds.

When we compare these results with our previous experimental data [22,23] of CaFe$_2$As$_2$ and Ca$_{0.6}$Na$_{0.4}$Fe$_2$As$_2$ compounds we find that in these compounds acoustic modes upto 12 meV do not show any change with partial doping of Na at the Ca site. The phonon modes from 12 to 40 meV in the superconducting Ca$_{0.6}$Na$_{0.4}$Fe$_2$As$_2$ are found to soften by about 1 to 2 meV in comparison with the parent CaFe$_2$As$_2$. The phonon softening in the superconducting Ca compound is attributed to the slightly longer Ca/Na-As and Fe-As bond lengths. It should be noted that in CaFe$_2$As$_2$ doping is at the Ca site, while the electronic bands near the Fermi level in CaFe$_2$As$_2$ arise mainly from the Fe and As atoms. So it is difficult to comment on the effect of change in electronic structure on the phonon modes of Ca$_{0.6}$Na$_{0.4}$Fe$_2$As$_2$. The situation in CaFeAsF is more favorable. We study the effect of Co doping at the Fe site on the phonon spectra. For CaFeAsF, Fe bands are mainly believed to be contributing to the superconductivity. We expect that Co doping at the Fe site would change the electronic density of states of the Fe$_{1-x}$Co$_x$ sub-lattice in the superconducting compounds in comparison with the parent CaFeAsF and this in turn could harden the acoustic and intermediate phonon spectra.

**B. Temperature dependence of phonon spectra for CaFe$_{1-x}$Co$_x$AsF compounds ($x$ = 0, 0.06, 0.12)**

Recently we have reported neutron powder diffraction experiments for CaFe$_{1-x}$Co$_x$AsF compounds ($x$ = 0, 0.06, 0.12). Our measurements show that the parent compound CaFeAsF undergoes a tetragonal to orthorhombic phase transition at 134 K followed by the magnetic transition at 114(3) K while cooling the sample. The long range antiferromagnetic order has been observed to coexist [15] with superconductivity in the orthorhombic phase of the under-doped CaFe$_{0.94}$Co$_{0.06}$AsF (onset T$_c$=20 K). The tetragonal to orthorhombic phase transition temperature for CaFe$_{0.94}$Co$_{0.06}$AsF is determined to be around 85(3) K, which is lower than that in the parent compound CaFeAsF. Magnetic order is found to be completely suppressed in optimally doped $CaFe_{0.88}Co_{0.12}AsF$ ($T_c$=20 K).

We have measured the temperature dependence of phonon spectra for the $CaFe_{1-x}Co_xAsF$ compounds ($x$ = 0, 0.06, 0.12) at 2 K, 125 K and 150 K. For $CaFe_{0.94}Co_{0.06}AsF$ we have collected one more data set at 50 K, since the tetragonal to orthorhombic phase transition in this compound is determined to be around 85(3) K while lowering the temperature. The temperature dependence of phonon spectra for $CaFe_{1-x}Co_xAsF$ compounds ($x$ = 0, 0.06, 0.12) is shown in Fig. 3. Phonon modes are expected to shift towards higher energies with decrease of the unit cell volume with decreasing temperature. However our measurements show that temperature variation across the tetragonal to orthorhombic phase transition or magnetic phase transition has little effect on the phonon spectra of the parent compound. It seems that the formation of Cooper pairs has no influence on the overall vibrational spectrum of $CaFe_{1-x}Co_xAsF$ compounds ($x$ = 0.06, 0.12) across the superconducting transition temperature (Tc=20 K). This observation is quite similar to our earlier studies on parent and superconducting $MFe_2As_2$ (M=Ba, Ca, Sr) compounds. The renormalization of specific phonons, if it exists, may however be difficult to detect in a powder sample.

The Bose factor corrected S(Q,E) plots for $CaFe_{1-x}Co_xAsF$ compounds ($x$ = 0, 0.06, 0.12) at various temperatures measured using the IN4C spectrometer are shown in Fig. 4. Our measurements do not show any clear signatures of spin or resonant spin excitations in parent and superconducting compounds respectively in the attainable (Q, E) range of IN4C. Recent inelastic measurements show evidence of spin [16(a)] and Resonant spin [16(b)] excitations in the antiferromagnetically and superconducting state of $BaFe_2As_2$ and $Ba_{0.6}K_{0.4}Fe_2As$, respectively. These measurements were carried out using the MERLIN spectrometer at ISIS. The (Q, E) range attainable at IN4C is different from that of MERLIN at low Q values. Further investigations might be necessary before drawing any final conclusions concerning $CaFe_{1-x}Co_xAsF$ compounds ($x$ = 0, 0.06, 0.12).

**C. Phonon calculations in $CaFe_{1-x}Co_xAsF$ compounds ($x$ = 0, 0.06, 0.12)**

The experimental phonon spectra collected for $CaFe_{1-x}Co_xAsF$ compounds ($x$ = 0, 0.06, 0.12) as a function of temperature are shown in Figs. 3. The phonon spectra have been calculated using both the shell model and first principles DFT-based methods. To simulate the non-stochiometry arising from the doping the shell model calculations are carried out for a $2 \times 2 \times 2$ super cell, where 6% and 12% of the Fe atoms are randomly replaced by Co atoms. The calculated phonon spectra (Fig. 1 and 3) for parent, 6% and 12% Co doped compounds show little difference as discussed in detail in Section IVA. However experimental data for 6% and 12% Co doped compounds show hardening of acoustic and intermediate energy modes in comparison with the parent compound.

In order to understand the contribution of various atomic motions to the phonon spectra we have calculated the partial densities of states (Fig. 2). The shell model calculations show that the Fe and As atoms mainly contribute in the 0–35 meV range, while the vibrations due to Ca and F atoms contribute in the entire 0–45 meV range, respectively. The Fe-As stretching modes are above 30 meV. The estimated range of vibrational frequencies for different atoms from ab-initio calculations is comparable with shell model calculations except for the Ca vibrations. DFT-based calculations show vibrations in the frequency range extending up to 35 meV for Ca atoms, comparable to the vibrational frequencies of Fe and As, whereas in the shell model calculation these vibrations extend up to 45 meV.

The ab-initio calculations (Fig. 3) are in excellent agreement with the inelastic neutron scattering data and empirical shell model calculations, when viewed with the resolution on IN4. It is worth noting that the differences between the magnetic and non-magnetic ab-initio calculations appear to be small with the magnetic calculation offering slightly better agreement with the experimental data. In order to understand these differences we have calculated partial densities of states for both cases. We find that vibrations frequencies of Ca and F atoms match very well in both the magnetic and nonmagnetic calculations (Fig. 2). However, there are slight differences in the vibrations of Fe and As in both cases. The inclusion of magnetic interactions results in a shift of the peak at 12 meV to lower energies in the Fe partial density of states. However, Fe modes above 23 meV are found to harden by 1 to 2 meV in comparison of the nonmagnetic calculations. In the case of As vibrations the magnetic ordering results in softening of low energy vibrations as for Fe atoms. This is most spectacular for modes at 25meV which are red-shifted by about 5 meV. The intermediate frequency vibrations of As atoms from 28 meV and above harden with the inclusion of magnetic ordering. The As atom vibrations are therefore a sensitive probe of the magnetic ordering in the Fe sub-lattice.

A more stringent check of the quality of the computed densities of states is provided by the higher resolution IN6 data in terms of spectral frequency, see Fig. 1. The relative intensities of low and high frequency peaks is not so well reproduced with respect to the IN6 data since the shorter wavelength on this instrument tends to lead to an incomplete sampling of reciprocal space and therefore the density of states. Fig. 1 shows the results of averaged magnetic interactions and precise magnetic ordering in the "broken symmetry" approach. The differences between the two kinds of magnetic calculation are minor but they both offer a significant improvement over the non-magnetic calculation, when these results are compared with the IN6 data. The similarity of the two types of magnetic calculations indicates that, although magnetic interactions have to be included in the calculation of interatomic forces, the details of magnetic ordering are less important. This is of relevance to the experimental observation (Fig. 3) that no significant changes are detected in the vibrational spectra when passing the magnetic phase transitions in these compounds.

These differences in the partial densities of states can also be understood from the phonon dispersion relations (Fig. 5). The comparison of the calculated dispersion relations in both the cases shows that for magnetic calculations the zone boundary modes around 12 meV are shifted to lower energies in comparison of nonmagnetic calculations. Similarly there is greater density of phonon branches in nonmagnetic calculations for energies around 23 to 25 meV, which results in new peak in the density of states at around 23 to 25 meV.

Another interesting feature to note is that the vibrational frequency range of all the atoms (Ca, Fe, As and F) in CaFeAsF is nearly the same. This is significantly different from our previous calculations [22, 23] on $CaFe_2As_2$ where Ca vibrations were limited to a narrow frequency range. This shows that bonding of the Ca atom is quite different in $CaFe_2As_2$ and CaFeAsF compounds due to additional Ca-F interactions and consequently the interaction between the Ca and Fe-As layer is also quite different.

## V. Conclusions

We have carried out systematic studies of the measurements of phonon densities of states in $CaFe_{1-x}Co_xAsF$ compounds ($x$ = 0, 0.06, 0.12). The comparison of the phonon spectra of superconducting and non-superconducting compounds shows pronounced differences. The phonon spectra of Co doped compounds strongly react to the amount of Co substitution in the parent compound. The renormalization of phonon modes is believed to be due to electron-phonon coupling. Our results then support coupling of electrons and phonons in Co doped CaFeAsF compounds. The calculated partial densities of states show that the range of Ca vibrations extend over the full energy range similar to Fe, As and F atoms. This indicates the existence of a quite different interaction between the Ca and the Fe-As layer, compared to that in $CaFe_2As_2$. Comparison of non-magnetic and magnetic, DFT-based calculations show that magnetic interactions have to be included in the Hessian of inter-atomic interactions in order to reproduce the measured GDOS but that the specific details of these interactions do not have a significant impact on the GDOS, as evidenced by the similarity of the "average magnetic" and " broken symmetry" models approaches.

FIG. 1 (Color online) Comparison of experimental phonon spectra for CaFe$_{1-x}$Co$_x$AsF ($x$ = 0, 0.06, 0.12). The phonon spectra are measured with incident neutron wavelength of 5.12 Å using the IN6 spectrometer at ILL. The calculated phonon spectra using the shell model and ab-initio are also shown. The calculated spectra have been convoluted with a Gaussian of FWHM of 5% of the energy transfer in order to describe the effect of energy resolution in the experiment.

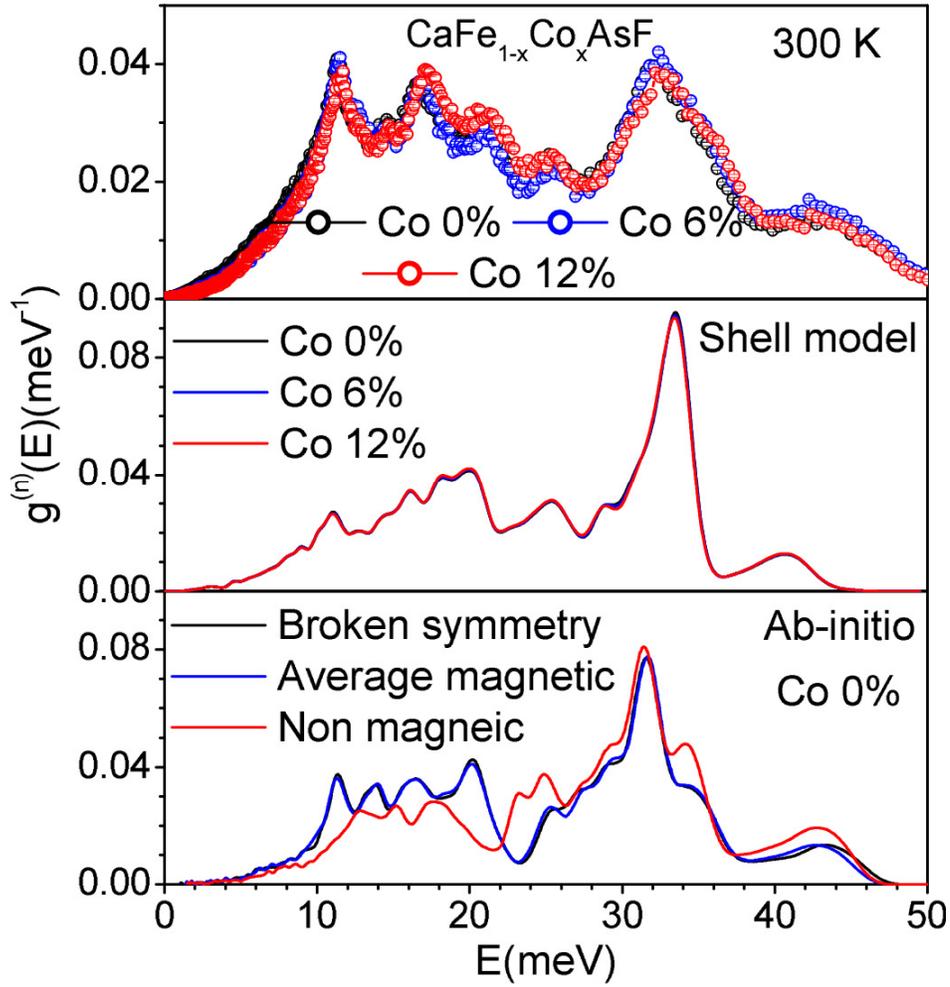

FIG. 2. (Color online) Calculated partial density of states for the various atoms in CaFe$_{1-x}$Co$_x$AsF ($x$ = 0, 0.06, 0.12) using the shell model and ab-inito calculations. In the right panel, "magnetic" refers to "average magnetic" calculations. The spectra are normalized to unity.

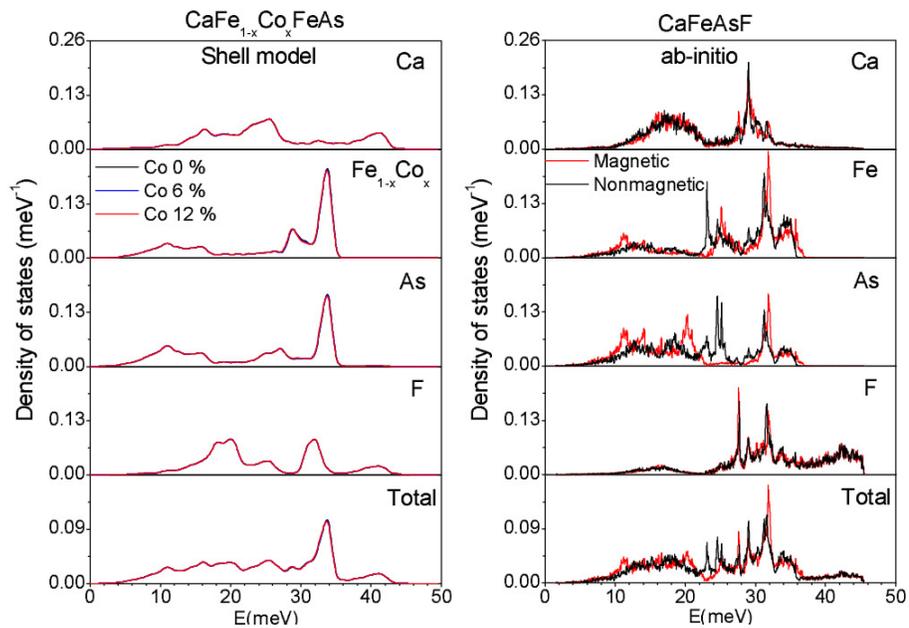

FIG. 3 (Color online) The temperature dependence of experimental phonon spectra for $CaFe_{1-x}Co_xAsF$ ($x$ = 0, 0.06, 0.12). The phonon spectra are measured with an incident neutron wavelength of 1.1 Å (67.6 meV) using the IN4C spectrometer at the ILL. The calculated phonon spectra using the shell model and ab-initio are also shown. In the bottom panel, "magnetic" refers to "average magnetic" calculations. The calculated spectra have been convoluted with a Gaussian of FWHM of 4 meV in order to describe the effect of energy resolution in the experiment.

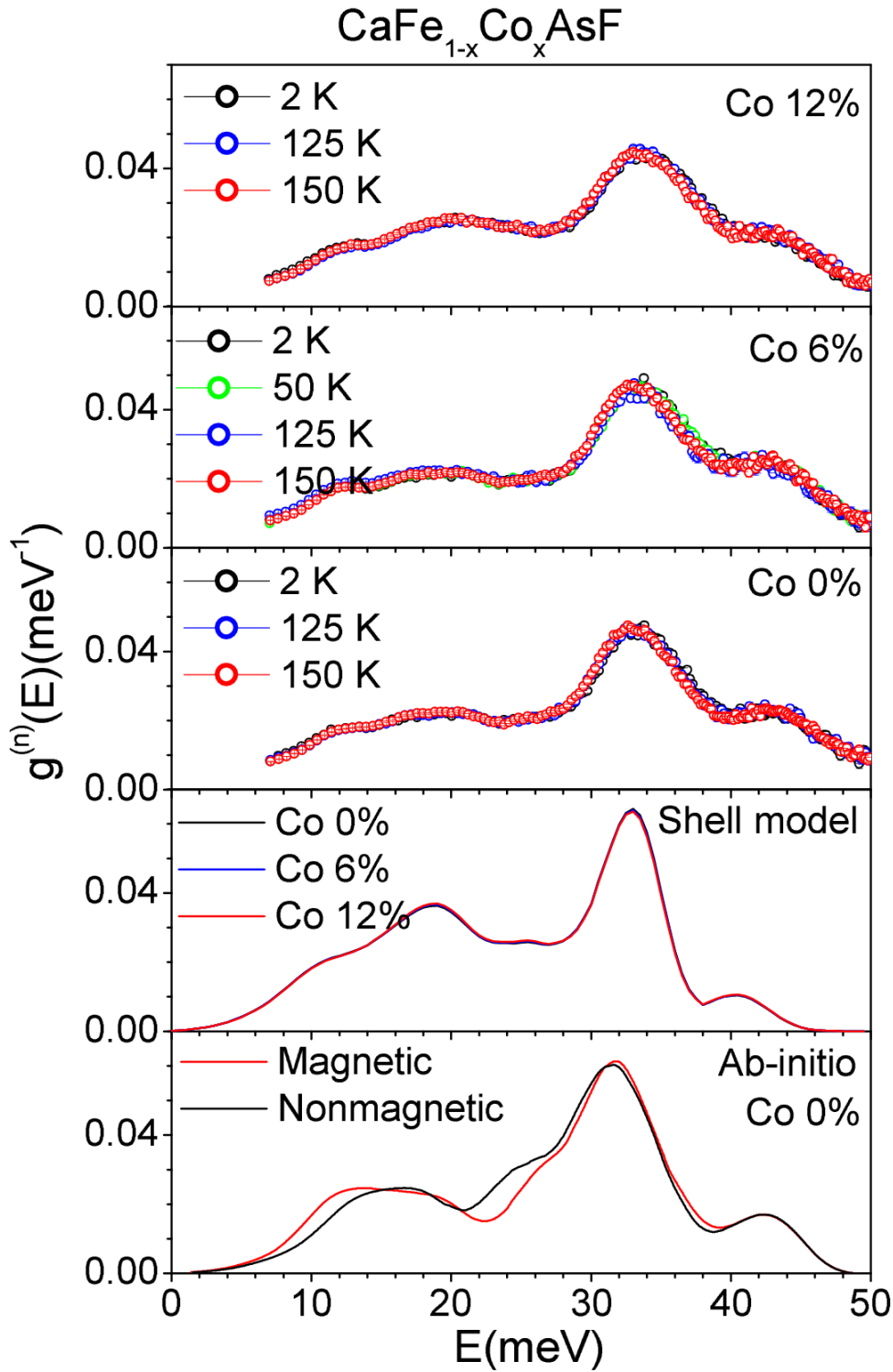

FIG. 4. (Color online) The experimental Bose factor corrected S(Q,E) plots for CaFe$_{1-x}$Co$_x$AsF ($x$ = 0, 0.06, 0.12) measured using the IN4C spectrometer at the ILL with an incident neutron wavelength of 1.1 Å. The values of S(Q,E) are normalized to the mass of sample in the beam. For clarity, a logarithmic representation is used for the intensities.

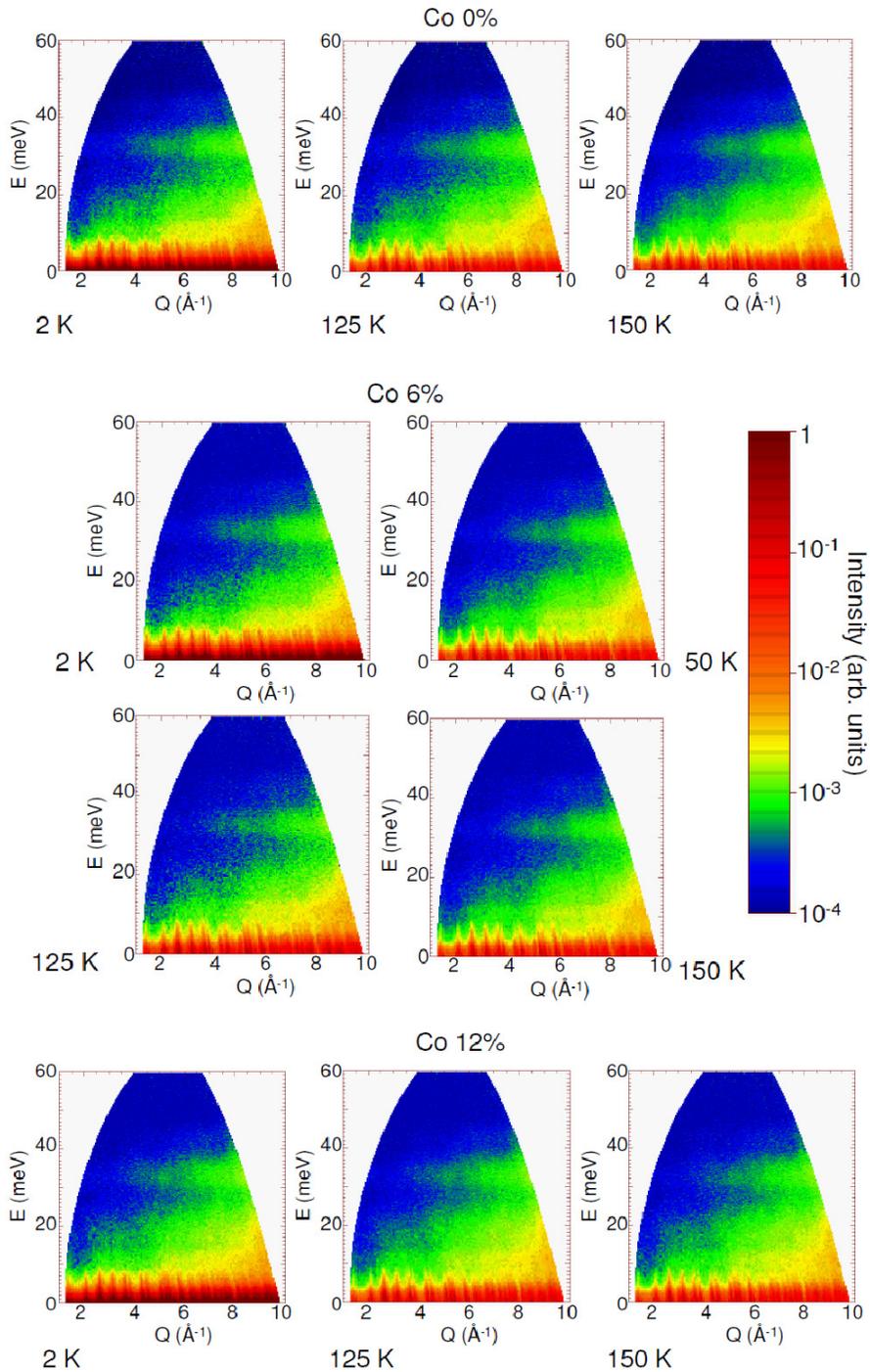

FIG. 5. Phonon dispersion relations for CaFeAsF from ab-initio calculations. In the upper panel, "magnetic" refers to "average magnetic" calculations. The Bradley-Cracknell notation is used for the high symmetry points along which the dispersion relations are obtained: Y= ( 12 , 1/2 , 0), Z=(0, 0, 1/2 ), T=( 1/2 , 1/2 , 1/2 ), S=(0, 1/2 , 0) and R=(0, 1/2 , 1/2 ).

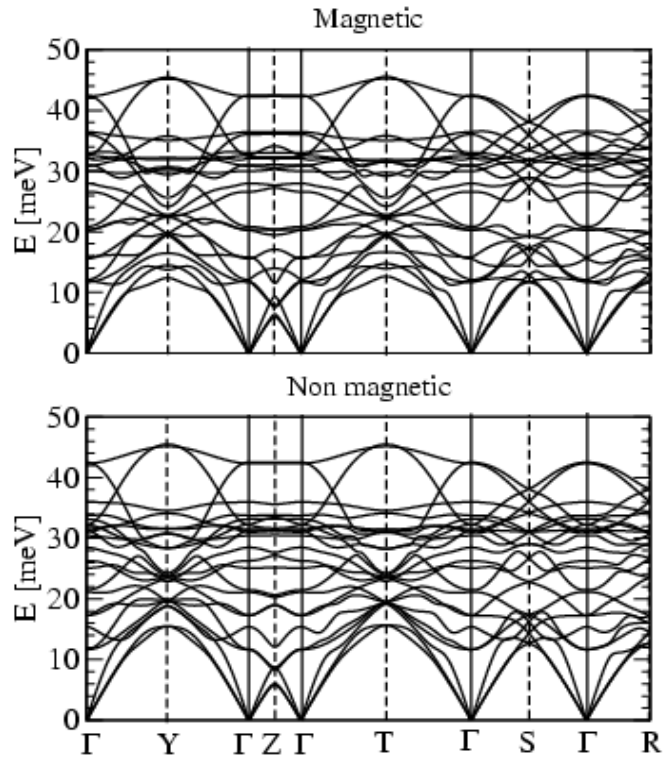